# 12-fold Symmetric Quasicrystallography from the lattices $F_4, B_6,$ and $E_6$

Nazife O. Koca[a)], Mehmet Koca[b)]
Department of Physics, College of Science, Sultan Qaboos University
P.O. Box 36, Al-Khoud, 123 Muscat, Sultanate of Oman
and
Ramazan Koc[c)]
Department of Physics, Gaziantep University, 27310, Gaziantep, Turkey

## ABSTRACT

One possible way to obtain the quasicrystallographic structure is the projection of the higher dimensional lattice into 2D or 3D subspaces. In this work we introduce a general technique applicable to any higher dimensional lattice. We point out that the Coxeter number and the integers of the Coxeter exponents of a Coxeter-Weyl group play a crucial role in determining the plane onto which the lattice to be projected. The quasicrystal structures display the dihedral symmetry of order twice the Coxeter number. The eigenvectors and the corresponding eigenvalues of the Cartan matrix are used to determine the set of orthonormal vectors in $n$D Euclidean space which lead suitable choices for the projection subspaces. The maximal dihedral subgroup of the Coxeter-Weyl group is identified to determine the symmetry of the quasicrystal structure. We give examples for 12-fold symmetric quasicrystal structures obtained by projecting the higher dimensional lattices determined by the affine Coxeter-Weyl groups $W_a(F_4)$, $W_a(B_6)$, and $W_a(E_6)$. These groups share the same Coxeter number $h=12$ with different Coxeter exponents. The dihedral subgroup $D_{12}$ of the Coxeter groups can be obtained by defining two generators $R_1$ and $R_2$ as the products of generators of the Coxeter-Weyl groups. The reflection generators $R_1$ and $R_2$ operate in the Coxeter planes where the Coxeter element $R_1R_2$ of the Coxeter-Weyl group represents the rotation of order 12. The canonical projections (strip projection, equivalently, cut and project technique) of the lattices determine the nature of the quasicrystallographic structures with 12-fold symmetry as well as the crystallographic structures with 4-fold and 6-fold symmetry. We note that the quasicrystal structures obtained from the lattices $W_a(F_4)$ and $W_a(B_6)$ are compatible with some experimental results.

[a)] electronic-mail: nazife@squ.edu.om
[b)] electronic-mail: kocam@squ.edu.om
[c)] electronic-mail: koc@gantep.edu.tr

# 1. Introduction

To describe the symmetry of a natural phenomenon the group theory plays a fundamental role. In this paper we will explore the quasicrystal structures obtained by projections of the lattices described by the Coxeter-Weyl groups. Applications of the Lie groups based on the Coxeter-Weyl groups are well known in High Energy Physics. The standard model of the High Energy Physics is based on the Lie group $SU(3) \times SU(2) \times U(1)$ (Weinberg, 1967; Salam, 1968; Fritzsch et al., 1973) and its extension to the grand unified theories $SU(5)$ (Georgi & Glashow, 1974), $SO(10)$ (Fritzsch & Minkowski, 1975), and $E_6$ (Gursey et al., 1976) are intimately related to their Coxeter-Weyl groups which are the point groups of the higher dimensional lattices described by the affine extensions. We wonder whether the Coxeter-Weyl groups $W(A_4)$, $W(D_5)$, and $W(E_6)$ of these Lie groups play any special role in the study of quasicrystallography. Any Coxeter-Weyl group with Coxeter number $h$ has a maximal dihedral subgroup of order $2h$ which acts faithfully in the Coxeter plane. The lattices described by the point groups $W(A_4)$ (Baake et al., 1990; Baake et al., 1990b; Koca et al., 2014) and $W(B_5)$ (de Brujin, 1981; Senechal, 1995) have been already proposed for the description of the quasicrystallography with 5-fold symmetries which result as the projections onto the respective Coxeter planes. The lattice described by the affine group $W_a(B_5)$ is the simple cubic lattice in 5D and may describe 10-fold symmetric quasicrystallography when projected onto a Coxeter plane. On the other hand, the lattices generated by the affine extension of the Coxeter-Weyl group $W(C_5)$ which has the same structure as the point group $W(B_5)$ are the root and weight lattices of the group $W_a(D_5)$. Note the isomorphism between the automorphism group of $D_5$ and the group $W(C_5)$, $Aut(D_5) \approx W(D_5):C_2 \approx W(C_5)$. This proves that the lattice $W_a(D_5)$ can be used to describe both 5-fold and 10-fold symmetric quasicrystallography. It was also observed that the quasicrystallographic structures in 3D with icosahedral symmetry can be described by canonical projections of the 6D lattices determined by the affine extension of the Coxeter-Weyl group $W(D_6)$ which embeds the non-crystallographic group $W(H_3)$ as a maximal subgroup (Kramer & Neri, 1984; Bak, 1986; Conway & Knowles, 1986; Shcherbak, 1988; Koca et al., 2001). In a recent paper (Koca et al., 2012) we have proposed that any quasicrystallographic structure with $h$-fold symmetry can be determined by projections of the higher dimensional lattices onto the relevant Coxeter planes.

After the discovery of the icosahedral quasicrystal by the D. Shechtman et al. (Shechtman et al., 1984), the recent developments indicate that one can compose metallic alloys or other chemical compositions which exhibit 5-fold, 8-fold, 10-fold, 12-fold, and 18-fold symmetries. For a general exposition see the references on quasicrystallography (Di Vincenzo & Steinhardt, 1991; Janot, 1989; Suck et al., 2010). A recent report on the quasicrystallography predicts the existence of a structure of 36-fold symmetry (Lubin et al., 2012). It seems there is no a priori any limitation on the order of the dihedral group which may describe the symmetry of the quasicrystal structures. This reminds us the classifications of the Coxeter-Weyl groups with different Coxeter numbers.

The Coxeter-Weyl groups are naturally characterized by some integers known as Coxeter exponents (Coxeter, 1952; Coxeter & Moser, 1965; Coxeter, 1973) and, in particular, each Coxeter-Weyl group has a Coxeter number $h$ which can be used to describe the dihedral subgroup $D_h$ of order $2h$. It is then quite natural to ask the question whether the projections of the higher dimensional lattices determined by the affine Coxeter-Weyl groups with Coxeter number $h$ can describe the quasicrystallography with $h$-fold symmetry. After introducing a general framework, we study the projections of the lattices described by the affine extensions of the Coxeter-Weyl groups $W_a(F_4)$, $W_a(B_6)$, and $W_a(E_6)$ whose Coxeter numbers are all given by $h=12$. The last two groups describe three different lattices in 6D Euclidean spaces and the first one defines a unique 4D lattice whose unit cell is the 24-cell. The Coxeter numbers of the Coxeter-Weyl groups $W(D_7)$ and $W(A_{11})$ are also equal to 12 but their lattice projections will not be included in this work.

There is yet another approach for the descriptions of the quasicrystal structures. This is a set theoretic approach initiated by Yves Meyer (Meyer, 1972; Meyer, 1995) and later developed by Robert V. Moody (Moody, 1995; Moody, 2000) in the name of *Model Set*. A set $\Lambda$ is a *Meyer set* if it is a *Delaunay* (*Delon*) *set* (Delaunay, 1932) and there exists a finite set *F* such that $\Lambda - \Lambda \subseteq \Lambda + F$, or equivalently, $\Lambda - \Lambda$ is uniformly discrete (Moody, 2000). J. C. Lagarias (Lagarias, 1996) prove that the Meyer set is also a Quasiperiodic set. In the context of the *Model set*, the cut and project techniques in the series of $E_8 \Rightarrow H_4, D_6 \Rightarrow H_3, A_4 \Rightarrow H_2$ is remarkable (Chen *et al.*, 1998) where the icosians ring plays an important role.

We organize the paper as follows. In Section 2 we study the general technique for the canonical projections of the lattices determined by the simply-laced Coxeter-Weyl groups $W(A_n)$, $W(D_n)$, $W(E_6)$, $W(E_7)$, and $W(E_8)$. The eigenvalues, eigenvectors and the simple roots of the Cartan matrix (Gram matrix) of the Coxeter-Weyl group are used for the determination of the Coxeter planes and its maximal dihedral subgroup. In Section 3 we study the canonical projection of the exceptional lattice described by the affine Coxeter-Weyl group $W_a(F_4)$. We project the fundamental polytopes of the group $W(F_4)$ onto the Coxeter plane, and in particular, point out the importance of the polytope which describe the Voronoi cell. We use the method of canonical projection (Duneau & Katz, 1985; Senechal, 1995) to obtain the 12-fold symmetric quasicrystal from the projection of the $W(F_4)$ lattice. It turns out that the quasicrystal structure obtained from $W(F_4)$ is compatible with an experimental observation (Ishimasa *et al.*, 1985). The Section 4 involves a general discussion of the lattice $Z^6$ described by the group $W_a(B_6)$ and the projections of its fundamental polytopes as well as the canonical projection of its 6D cubic lattice onto the Coxeter plane. The quasicrystal structure obtained by projection agrees with the experimental result published recently (Forster *et al.*, 2013). In Section 5 we apply the similar technique developed in Section 2 to the root and weight lattices described by the group $W_a(E_6)$. Section 6 is devoted for the discussions of the relevant techniques and further extensions of the work.

## 2. Decomposition of the lattice spaces into orthogonal planes and the canonical projection

Let *G* be the Coxeter-Dynkin diagram (Coxeter, 1973) representing the Coxeter-Weyl group of rank *n* which is described by the simple roots $\alpha_i, (i=1,2,...,n)$. In this section we will only consider the simply-laced root systems with the norm fixed by $(\alpha_i, \alpha_i) = 2$. The Cartan matrix representing non-orthogonality of the simple roots is defined by the scalar product

$$C_{ij} = \frac{2(\alpha_i, \alpha_j)}{(\alpha_j, \alpha_j)}. \tag{1}$$

It is a real and symmetric matrix for the simply-laced root systems and also called the Gram matrix by the crystallographers. The weight vectors $\omega_i$ are defined by the relation $(\omega_i, \alpha_j) = \delta_{ij}$ where $\delta_{ij}$ is the Kronocker-delta. Weight vectors and the roots are related to each other by the relations

$$\alpha_i = \sum_j C_{ij} \omega_j, \quad \omega_i = \sum_j (C^{-1})_{ij} \alpha_j. \tag{2}$$

The root lattice $G$ and the weight lattice $G^*$ are defined respectively by the vectors $p = \sum_{i=1}^n b_i \alpha_i$, $b_i \in \mathbf{Z}$ and $q = \sum_{i=1}^n a_i \omega_i$, $a_i \in \mathbf{Z}$. A standard notation for the weight vector is $q \equiv (a_1, a_2, ..., a_n)$, with $a_i \in \mathbf{Z}$. Let $r_i, (i = 1, 2, ..., n)$ be the reflection generator with respect to the hyperplane orthogonal to the simple root $\alpha_i$. We will consider the highest weight vector $(a_1, a_2, ..., a_n)$ with $a_i \geq 0$, which designates the irreducible representation of a Lie group associated with the Coxeter-Weyl group (Slansky, 1981). Denote by $W(G) = \langle r_1, r_2, ..., r_n | (r_i r_j)^{m_{ij}} = 1 \rangle$ the Coxeter-Weyl group generated by the reflections. The affine extension of the Coxeter-Weyl group $W_a(G)$ can be generated by adjoining a generator $r_0$ which represents the reflection with respect to the hyperplane bisecting the highest root which can be determined from the extended Coxeter-Dynkin diagram of the Lie group. We use the notation $W(G)t \equiv t_G$ for the orbit generated from the highest weight vector $t$ by the Coxeter group $W(G)$. The set of vectors $t_G = (a_1, a_2, ..., a_n)_G$ describes the vertices of a polytope and should not be confused with the vector $(a_1, a_2, ..., a_n)$, with $a_i \geq 0$ used for the designation of the irreducible representation of the associated Lie algebra. We assume that the set of simple roots can be decomposed into two sets in such a way that the corresponding reflection generators $r_1, r_2, ..., r_k$ commute pairwise as well as do the set $r_{k+1}, r_{k+2}, ..., r_n$ for some $k < n$. Define the generators $R_1$ and $R_2$ by $R_1 = r_1 r_2 ... r_k$ and $R_2 = r_{k+1} r_{k+2} ... r_n$ (Carter, 1972; Humphreys, 1990). It is easy to show that the generators $R_1$ and $R_2$ act as reflections on the simple roots of the Coxeter diagram $I_2(h)$. The Coxeter element $R_1 R_2$ represents a rotation of order $h$ in the plane spanned by the simple roots of the graph $I_2(h)$. The integers $m_i$ of the Coxeter exponents $m_i \frac{\pi}{h} (i = 1, 2, ..., n)$ (Carter, 1972; Humphreys, 1990) listed in Table 1 are important for the determinations of the orthogonal planes in which the Coxeter element acts as a rotation. The eigenvalues of the Cartan matrix can be written simply as $2[1 - \cos(m_i \frac{\pi}{h})]$. The $D_n$ series with even $n$ has two degenerate eigenvalues corresponding to the eigenvalue 2. The Cartan matrix of the Coxeter-Weyl group of rank odd $n$ has an eigenvalue 2. The lattice of even dimension $n$ described by the affine Coxeter group of rank-$n$ can be decomposed into orthogonal planes determined by the pair of orthogonal unit vectors obtained from the eigenvectors of the Cartan matrix corresponding to the eigenvalues $2[1 - \cos(m_i \frac{\pi}{h})]$ and $2[1 - \cos(h - m_i) \frac{\pi}{h}]$.

Table1. Integers of the Coxeter exponents and the Coxeter number

|   | $m_1, m_2, ..., m_n$ | $h$ |
|---|---|---|
| $A_n$ | $1, 2, ..., n$ | $n+1$ |
| $B_n$ | $1, 3, 5, ..., 2n-1$ | $2n$ |
| $C_n$ | $1, 3, 5, ..., 2n-1$ | $2n$ |
| $D_n$ | $1, 3, 5, ..., 2n-3, n-1$ | $2(n-1)$ |
| $E_6$ | $1, 4, 5, 7, 8, 11$ | $12$ |
| $E_7$ | $1, 5, 7, 9, 11, 13, 17$ | $18$ |
| $E_8$ | $1, 7, 11, 13, 17, 19, 23, 29$ | $30$ |
| $G_2$ | $1, 5$ | $6$ |
| $F_4$ | $1, 5, 7, 11$ | $12$ |
| $H_3$ | $1, 5, 9$ | $10$ |
| $H_4$ | $1, 11, 19, 29$ | $30$ |
| $I_2(h)$ | $1, h-1$ | $h$ |

However in the case of odd $n$, the eigenvector corresponding to the eigenvalue 2 singles out. Let $\lambda_i, (i=1,2,...,n)$ be the eigenvalue and the $\vec{X}_i$ is the corresponding normalized eigenvector satisfying the eigenvalue equation

$$\sum_j C_{ij} X_{jk} = \lambda_k X_{ik} . \qquad (3)$$

Since the Cartan matrix of a simply-laced Coxeter-Weyl group is real and symmetric it follows that the eigenvalues are real and the corresponding eigenvectors are orthogonal. However, care should be taken for the Coxeter-Weyl group $D_n$ for even $n$ since the eigenvalue 2 is doubly degenerate. The above equation then can be written as

$$(X^T C X)_{ij} = \lambda_i \delta_{ij} . \qquad (4)$$

This will be used for the simply laced group $W(E_6)$. One can define the orthogonal set of unit vectors $\hat{x}_i$ by the relation

$$\hat{x}_i = \frac{1}{\sqrt{\lambda_i}} \sum_j \alpha_j X_{ji} . \qquad (5)$$

It is straightforward to prove that the vectors in (5) form an orthonormal set spanning an $n$-dimensional Euclidean space. A general proof can be obtained for all Coxeter groups by simple modification of the proof given in reference (Steinberg, 1951; Carter, 1972). Consequently, a similar relation for the non-simply laced groups can be obtained where $\hat{x}_i = \frac{1}{\sqrt{h}\sqrt{\lambda_i}} \sum_j X_{ji} \frac{2\alpha_j}{(\alpha_j, \alpha_j)}$. Here the eigenvectors are such that the last components are all equal 1.

Let us further elaborate the properties of the planes in which the dihedral group $D_h$ (it is unfortunate that the same letter is also used for the Coxeter-Weyl group) or some of its subgroups. We have already stated that the dihedral group acts in the plane determined by the unit vectors corresponding to the eigenvalues $2[1-\cos(m_i\frac{\pi}{h})]$ and $2[1-\cos(h-m_i)\frac{\pi}{h}]$. For odd $n$ one of the integers takes the value $m_i = \frac{h}{2}$ so that the corresponding eigenvalue is 2. The corresponding unit vector to this eigenvalue is unpaired with any other unit vector. Therefore it will be orthogonal to the rest of the planes determined by the other unit vectors. If we choose $m_1 = 1$ and $m_n = h-1$ (in the case of the Coxeter-Weyl group $D_n$ this reads $m_1 = 1$ and $m_{n-1} = h-1$) then the corresponding eigenvalues will be $2[1 \mp \cos(\frac{\pi}{h})]$. Let us denote the corresponding unit vectors by $\hat{x}_1$ and $\hat{x}_n$ ($\hat{x}_{n-1}$ for $D_n$). Then one can show that the root vectors

$$\beta_1 = \sqrt{2}[\sin(\frac{\pi}{2h})\hat{x}_1 + \cos(\frac{\pi}{2h})\hat{x}_n], \quad \beta_n = \sqrt{2}[\sin(\frac{\pi}{2h})\hat{x}_1 - \cos(\frac{\pi}{2h})\hat{x}_n] \qquad (6)$$

determine the Coxeter plane and the simple roots of the Coxeter graph $I_2(h)$ satisfy the Cartan matrix

$$\begin{pmatrix} 2 & -2\cos(\frac{\pi}{h}) \\ -2\cos(\frac{\pi}{h}) & 2 \end{pmatrix}. \qquad (7)$$

It is easy to prove that the generators $R_1$ and $R_2$ act as reflections on the simple roots $\beta_1$ and $\beta_n$ respectively. Consequently, the Coxeter element $R_1 R_2$ acts as a rotation by $\frac{2\pi}{h}$ in the Coxeter plane determined by the vectors $\beta_1$ and $\beta_n$. All orthogonal planes can be compactly defined by the pair of simple roots $\beta_i, \beta_{n+1-i}$ with

$$\beta_i = \sqrt{2}[\sin(\frac{m_i \pi}{2h})\hat{x}_i + \cos(\frac{m_i \pi}{2h})\hat{x}_{n+1-i}],$$
$$\beta_{n+1-i} = \sqrt{2}[\sin(\frac{m_i \pi}{2h})\hat{x}_i - \cos(\frac{m_i \pi}{2h})\hat{x}_{n+1-i}], \; (i=1,2,...,\frac{n}{2}) \qquad (8)$$

for even $n$. Therefore each pair of simple roots $\beta_i, \beta_{n+1-i}$ determines a Coxeter graph $I_2(\frac{h}{m_i})$ where the Coxeter element acts as a rotation by the angle $m_i \frac{2\pi}{h}$. The scalar product $(\beta_i, \beta_{n+1-i}) = -2\cos(\frac{m_i \pi}{h})$ implies that the lattice space is decomposed into orthogonal set of planes each of which is associated with a pair of Coxeter exponents satisfying $m_i + m_{n+1-i} = h$. Similar decomposition of the lattice space into orthogonal set of planes was studied by Barache et al., (Barache et al., 1998). Their main concern was the Pisot-cyclotomic numbers which are used to determine the planes $\beta_i, \beta_{n+1-i}$ ($u_i, v_i$ are their notation for the planes) and did not refer to the Coxeter exponents derived from the eigenvalues of the Cartan matrix and the eigenvectors used to define the set of orthogonal vectors given in (5). The Cartan matrix of an arbitrary Coxeter-Weyl group can be transformed into a block-diagonal matrix where each entry is represented by a $2\times 2$ matrix similar to (7) for the even $n$ otherwise it includes a $1\times 1$ matrix with entry 2. Similarly, the Coxeter element can be represented by a block-diagonal matrix

$$R_1 R_2 = \begin{pmatrix} M_1 & 0 & 0 & 0 & 0 & 0 \\ 0 & M_2 & 0 & 0 & 0 & 0 \\ 0 & 0 & . & 0 & 0 & 0 \\ 0 & 0 & 0 & . & 0 & 0 \\ 0 & 0 & 0 & 0 & . & 0 \\ 0 & 0 & 0 & 0 & 0 & M_s \end{pmatrix} \qquad (9)$$

where each $M_s$, acting on the set of orthogonal unit vectors $\hat{x}_i$, is either $1, -1$ or a $2\times 2$ matrix of the form ( Engel, 1986; Senechal, 1995)

$$\begin{pmatrix} \cos(m_i \frac{2\pi}{h}) & -\sin(m_i \frac{2\pi}{h}) \\ \sin(m_i \frac{2\pi}{h}) & \cos(m_i \frac{2\pi}{h}) \end{pmatrix}. \qquad (10)$$

The planes determined by (10) are called the principal planes by Engel (Engel, 1986). The characteristic equation of the Coxeter element is the product of the characteristic equations obtained from (10). A similar result was obtained from a different approach in Section 5.3 of the reference (Engel, 1986) for an arbitrary lattice. Components of the root and weight vectors $p = \sum_{i=1}^{n} b_i \alpha_i$ and $q = \sum_{i=1}^{n} a_i \omega_i \equiv (a_1, a_2, ..., a_n)$ in the orthonormal basis can be easily determined by using (5) as

$$p_i = \sqrt{\lambda_i} \sum_{j=1}^{n} b_j X_{ji}, \qquad q_i = \frac{1}{\sqrt{\lambda_i}} \sum_{j=1}^{n} a_j X_{ji}; \quad a_i, b_i \in \mathbf{Z} . \qquad (11)$$

The pair of components $(p_i, p_{n+1-i})$ and $(q_i, q_{n+1-i})$ in the plane determined by the root and weight vectors represent respectively the orthogonal projections of the respective lattices. The projection technique given in (11) has not been discussed in this context elsewhere. An $n$-dimensional Euclidean space can be decomposed as the space $E_{\parallel}$ determined by the pair of unit vectors $\hat{x}_1$ and $\hat{x}_n$ and the complement is the orthogonal space $E_{\perp}$. The canonical projection can be defined as follows. Pick up the set of vectors $V(0)$ (MacKay, 1982; Katz, 1988; Katz, 1989). that represents the Voronoi cell of the root (Conway & Sloane, 1982; Conway & Sloane, 1988); Moody & Patera, 1992) or weight lattices (Conway & Sloane, 1988) around the origin. Projection of the Voronoi cell into the space $E_{\perp}$ determines a ball whose surface is a sphere $S^{n-3}$ of radius $R_0$ which is also called window $\mathbf{K}$ (Senechal, 1995). Then one projects the lattice points into the ball determined by the radius $R_0$. Then project those lattice points which are restricted by the sphere of radius $R_0$ onto the plane determined by the unit vectors $\hat{x}_1$ and $\hat{x}_n$. This is a general technique for every higher dimensional lattice determined by the affine Coxeter groups. In the following sections we apply this technique for the projections of the lattices determined by the Coxeter-Weyl groups $W_a(F_4)$, $W_a(B_6)$, and $W_a(E_6)$.

A few words are in order for the Voronoi cells of the lattices. The Voronoi regions of the root lattices are discussed in detail in reference (Moody & Patera, 1992) and can be easily summarized in our notations as follows. The Voronoi cell $V(0)$ of $A_n$ is a polytope dual to the root polytope $(1,0,0,...,0,1)_{A_n}$ and can be written as the union of the orbits $(1,0,0,...,0,0)_{A_n} \cup (0,1,0,...,0,0)_{A_n} \cup ... \cup (0,0,0,...,0,1)_{A_n}$. For instance the Voronoi cell of the fcc lattice of $A_3$ (the root lattice) is a Catalan solid, the rhombic dodecahedron, described by the union of the polytopes $(1,0,0)_{A_3} \cup (0,1,0)_{A_3} \cup (0,0,1)_{A_3}$ dual to the root polytope $(1,0,1)_{A_3}$ (Koca *et al.*, 2010). Similarly the Voronoi cell of $A_4$ is the union of the polytopes $(1,0,0,0)_{A_4} \cup (0,1,0,0)_{A_4} \cup (0,0,1,0)_{A_4} \cup (0,0,0,1)_{A_4}$ dual to the root polytope $(1,0,0,1)_{A_4}$ (Koca *et al.*, 2012). One can check that for all simply laced Coxeter-Weyl groups the Voronoi cell of the root lattice is the dual polytope of the root polytope. The Voronoi cells of the weight lattices have been studied in reference (Conway & Sloane, 1988) that can be expressed in terms of our notation. The Voronoi cell $V(0)$ of the weight lattice $A_n^*$ (Koca *et al.*, 2012) can be represented by the polytope $\frac{1}{n+1}(1,1,...,1)_{A_n}$.

## 3. Canonical projection of the lattice generated by the exceptional group $W_a(F_4)$

The Coxeter-Dynkin diagram of the exceptional group $W(F_4)$ is shown in Figure 1. The order of the group is $|W(F_4)| = 1152$.

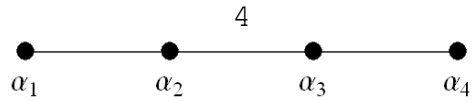

Figure 1. The Coxeter-Dynkin diagram of $W(F_4)$

It has two long roots (first two from left) and two short roots (last two). The Cartan matrix and its inverse are given as follows

$$C_{F_4} = \begin{pmatrix} 2 & -1 & 0 & 0 \\ -1 & 2 & -2 & 0 \\ 0 & -1 & 2 & -1 \\ 0 & 0 & -1 & 2 \end{pmatrix}, \quad (C_{F_4})^{-1} = \begin{pmatrix} 2 & 3 & 4 & 2 \\ 3 & 6 & 8 & 4 \\ 2 & 4 & 6 & 3 \\ 1 & 2 & 3 & 2 \end{pmatrix}. \qquad (12)$$

The dual space basis vectors are defined by the relation $(\omega_i, \frac{2\alpha_j}{(\alpha_j, \alpha_j)}) = \delta_{ij}$ where the scalar product in the dual space is defined by the metric tensor $(\omega_i, \omega_j) = G_{ij}$ which is given in the matrix form as

$$G = \begin{pmatrix} 2 & 3 & 2 & 1 \\ 3 & 6 & 4 & 2 \\ 2 & 4 & 3 & \frac{3}{2} \\ 1 & 2 & \frac{3}{2} & 1 \end{pmatrix}. \tag{13}$$

The $F_4$ lattice is generated by its short roots and evidently the lattice is self dual because the inverse matrix in (12) has integer entries only. Therefore the root and the weight lattices coincide up to an action of the Coxeter-Weyl group $W_a(F_4)$. We will work with the dual space basis vectors to define a general lattice vector $q = \sum_{i=1}^{4} a_i \omega_i \equiv (a_1, a_2, a_3, a_4)$, $a_i \in \mathbf{Z}$. In an earlier paper (Koca *et al.*, 2013) the $F_4$ polytopes have been studied in detail. The self-dual regular polytope of $F_4$ is called the 24-cell and it is either represented by the orbit $(1,0,0,0)_{F_4}$ or $(0,0,0,1)_{F_4}$. In Figure 2 we display the orthogonal projections of some of the $W(F_4)$ polytopes.

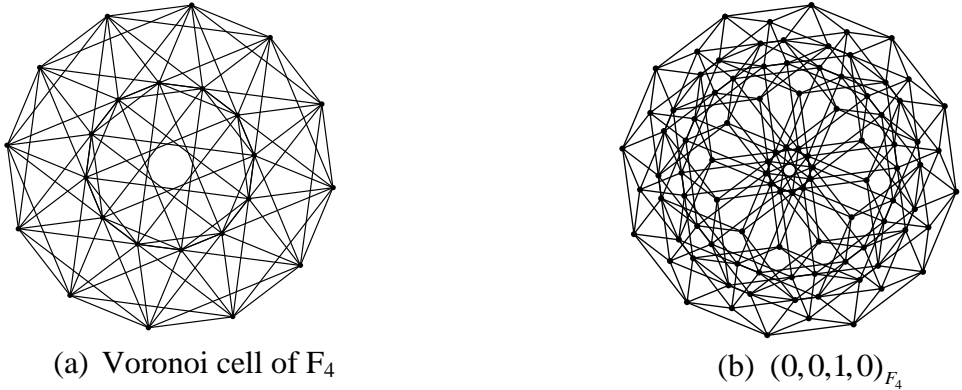

(a) Voronoi cell of $F_4$   (b) $(0,0,1,0)_{F_4}$

Figure 2. Orthogonal projections of some of the $W(F_4)$ polytopes

The only difference between these two representations of the 24-cell is that norms of the vectors are given by $(\omega_1, \omega_1) = 2$ versus $(\omega_4, \omega_4) = 1$ and one is rotated with respect to the other within the lattice space. A quaternionic representation of the polytope $(1,0,0,0)_{F_4}$ corresponds to the quaternionic elements of the binary tetrahedral group $T$ (Koca *et al.*, 2006). The $F_4$ lattice is generated by the polytope $(0,0,0,1)_{F_4}$ and the scaled copy of the 24-cell $\frac{1}{2}(0,0,0,1)_{F_4}$ (Moody & Patera, 1992) represents the Voronoi cell of the lattice. The four orthonormal unit vectors can be constructed by a similar technique studied in the previous section. The lattice space of $F_4$ can be decomposed into two orthogonal planes represented by the pair of vectors $(\beta_1, \beta_4)$ and $(\beta_2, \beta_3)$ which are given by

$$\beta_1 = \sqrt{2}[\sin(\frac{\pi}{24})\hat{x}_1 + \cos(\frac{\pi}{24})\hat{x}_4] = \frac{1}{\sqrt{3+\sqrt{3}}}[\alpha_1 + (\sqrt{3}+1)\alpha_3],$$

$$\beta_4 = \sqrt{2}[\sin(\frac{\pi}{24})\hat{x}_1 - \cos(\frac{\pi}{24})\hat{x}_4] = \frac{1}{\sqrt{3+\sqrt{3}}}[\frac{(\sqrt{3}+1)}{\sqrt{2}}\alpha_2 + \sqrt{2}\alpha_4],$$

$$\beta_2 = \sqrt{2}[\sin(\frac{5\pi}{24})\hat{x}_2 + \cos(\frac{5\pi}{24})\hat{x}_3] = \frac{1}{\sqrt{3-\sqrt{3}}}[\alpha_1 + (1-\sqrt{3})\alpha_3],$$

$$\beta_3 = \sqrt{2}[\sin(\frac{5\pi}{24})\hat{x}_2 - \cos(\frac{5\pi}{24})\hat{x}_3] = \frac{1}{\sqrt{3-\sqrt{3}}}[\frac{(1-\sqrt{3})}{\sqrt{2}}\alpha_2 + \sqrt{2}\alpha_4].$$

(14)

The Coxeter element represents a rotation by $30°$ in the plane $(\beta_1, \beta_4)$ and $150°$ in the plane $(\beta_2, \beta_3)$. The orthogonal components of a general lattice vector in the plane $(\hat{x}_1, \hat{x}_4)$ can be written as

$$q_1 = \frac{1}{\sqrt{2}\sqrt{2-\sqrt{2+\sqrt{3}}}\sqrt{3+\sqrt{3}}}[a_1 + \frac{\sqrt{3}+1}{\sqrt{2}}a_2 + \frac{\sqrt{3}+1}{2}a_3 + \frac{1}{\sqrt{2}}a_4],$$

$$q_4 = \frac{1}{\sqrt{2}\sqrt{2+\sqrt{2+\sqrt{3}}}\sqrt{3+\sqrt{3}}}[a_1 - \frac{\sqrt{3}+1}{\sqrt{2}}a_2 + \frac{\sqrt{3}+1}{2}a_3 - \frac{1}{\sqrt{2}}a_4],$$

(15)

and in the plane $(\hat{x}_2, \hat{x}_3)$ as

$$q_2 = \frac{1}{\sqrt{2}\sqrt{2-\sqrt{2-\sqrt{3}}}\sqrt{3-\sqrt{3}}}[a_1 + \frac{1-\sqrt{3}}{\sqrt{2}}a_2 + \frac{1-\sqrt{3}}{2}a_3 + \frac{1}{\sqrt{2}}a_4],$$

$$q_3 = \frac{1}{\sqrt{2}\sqrt{2+\sqrt{2-\sqrt{3}}}\sqrt{3-\sqrt{3}}}[a_1 - \frac{1-\sqrt{3}}{\sqrt{2}}a_2 + \frac{1-\sqrt{3}}{2}a_3 - \frac{1}{\sqrt{2}}a_4].$$

(16)

When the Voronoi cell $\frac{1}{2}(0,0,0,1)_{F_4}$ is projected onto the plane $(\hat{x}_2, \hat{x}_3)$ it describes a disc of radius $R_0 = \frac{1}{2}(q_2^2 + q_3^2)^{\frac{1}{2}}$. One can select the lattice points which project into this region with their components $q_2$ and $q_3$ satisfying the relation $\frac{1}{2}(q_2^2 + q_3^2)^{\frac{1}{2}} \leq R_0$. Now project those points onto the plane $(\beta_1, \beta_4)$. The result of this projection is shown in Figure 3. One can repeat the same procedure by interchanging the planes that leads to the same result since the dihedral group acts in both planes faithfully. The quasicrystal structure in Figure 3 is strikingly similar observed in the quasicrystals of the Ni-Cr particles (Ishimasa *et al.*, 1985).

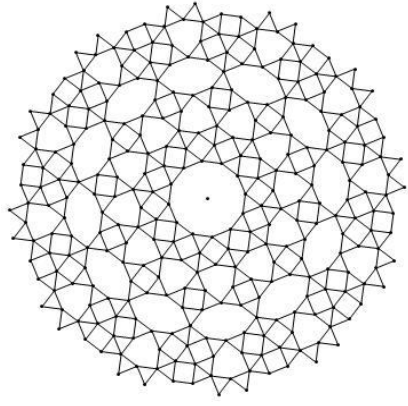

Figure 3. The canonical projection of the $F_4$ lattice onto the plane $(\hat{x}_1, \hat{x}_4)$

**4. Canonical projection of the lattice generated by the group $W_a(B_6)$**

The Coxeter-Dynkin diagram of the group $W(B_6)$ is displayed in Figure 4. The order of the group is $|W(B_6)| = 2^6 6!$.

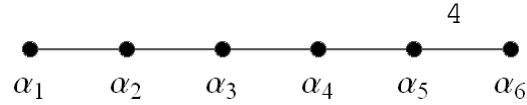

Figure 4. The Coxeter-Dynkin diagram of the Coxeter-Weyl group $W(B_6)$

Its last root is a short root of norm 1. The Cartan matrix and its inverse are given as follows:

$$C_{B_6} = \begin{pmatrix} 2 & -1 & 0 & 0 & 0 & 0 \\ -1 & 2 & -1 & 0 & 0 & 0 \\ 0 & -1 & 2 & -1 & 0 & 0 \\ 0 & 0 & -1 & 2 & -1 & 0 \\ 0 & 0 & 0 & -1 & 2 & -2 \\ 0 & 0 & 0 & 0 & -1 & 2 \end{pmatrix}, \quad (C_{B_6})^{-1} = \begin{pmatrix} 1 & 1 & 1 & 1 & 1 & 1 \\ 1 & 2 & 2 & 2 & 2 & 2 \\ 1 & 2 & 3 & 3 & 3 & 3 \\ 1 & 2 & 3 & 4 & 4 & 4 \\ 1 & 2 & 3 & 4 & 5 & 5 \\ \frac{1}{2} & 1 & \frac{3}{2} & 2 & \frac{5}{2} & 3 \end{pmatrix}. \quad (17)$$

The metric tensor in the dual space is given by the matrix $(\omega_i, \omega_j) = G_{ij} = (C^{-1})_{ij} \dfrac{(\alpha_j, \alpha_j)}{2}$.

The root lattice is a simple cubic lattice in 6D generated by the short roots of $B_6$. The lattice can be represented in the weight space with the lattice vectors given by

$q = \sum_{i=1}^{5} a_i \omega_i + 2a_6 \equiv (a_1, a_2, ..., 2a_6)$, $a_i \in \mathbf{Z}$. For example, the Voronoi cell is the polytope $(0,0,0,0,0,1)_{B_6}$ which represents a 6D cube with 64 vertices. Its fundamental polytopes are represented by the orbit $(1,0,0,0,0,0)_{B_6}, (0,1,0,0,0,0)_{B_6}, (0,0,1,0,0,0)_{B_6}, (0,0,0,1,0,0)_{B_6}$, $(0,0,0,0,1,0)_{B_6}$, and $(0,0,0,0,0,1)_{B_6}$. For example, the two polytopes $(1,0,0,0,0,0)_{B_6}$ and $(0,0,0,0,0,1)_{B_6}$ are the dual to each other. Their properties are described in Table 2. The vertices of the 6D cube can also be written as $(0,0,0,0,0,1)_{B_6} = \frac{1}{2}(\pm l_1 \pm l_2 \pm l_3 \pm l_4 \pm l_5 \pm l_6)$ in terms of some suitable orthonormal set of vectors with $(l_i, l_j) = \delta_{ij}, (i, j = 1, 2, ..., 6)$ where the simple roots are defined as $\alpha_1 = l_1 - l_2, \alpha_2 = l_2 - l_3, \alpha_3 = l_3 - l_4, \alpha_4 = l_4 - l_5, \alpha_5 = l_5 - l_6, \alpha_6 = l_6$.

Table 2. Numbers of facets of some of the $B_6$ polytopes

| Polytope | $N_0$ | $N_1$ | $N_2$ | $N_3$ | $N_4$ | $N_5$ |
|---|---|---|---|---|---|---|
| $(1,0,0,0,0,0)_{B_6}$ | 12 | 60 | 160 | 240 | 192 | 64 |
| $(0,1,0,0,0,0)_{B_6}$ | 60 | 480 | 1,120 | 1,200 | 576 | 76 |
| $(0,0,1,0,0,0)_{B_6}$ | 160 | 1,440 | 2,880 | 2,160 | 636 | 76 |
| $(0,0,0,1,0,0)_{B_6}$ | 240 | 1,920 | 3,200 | 2,080 | 636 | 76 |
| $(0,0,0,0,1,0)_{B_6}$ | 192 | 960 | 1,600 | 1,200 | 444 | 76 |
| $(0,0,0,0,0,1)_{B_6}$ | 64 | 192 | 240 | 160 | 60 | 12 |

The numbers $N_0 - N_1 + N_2 - N_3 + N_4 - N_5 = 0$ satisfy the Euler equation.

The projections of some of those polytopes onto the Coxeter plane are depicted in Figure 5.

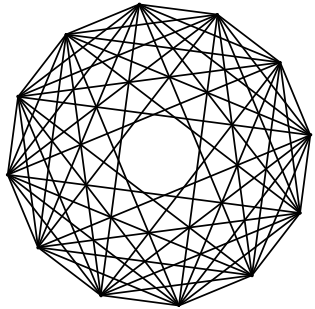 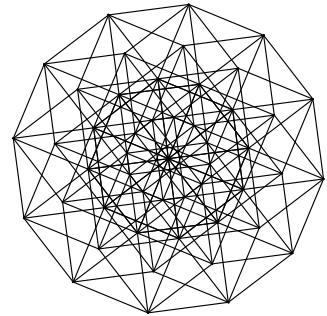

$(1,0,0,0,0,0)_{B_6}$ : Octahedron in 6D  $\quad\quad\quad$  $(0,0,0,0,0,1)_{B_6}$ : Cube in 6D

Figure 5. Orthogonal projections of some of the polytopes of $W(B_6)$

The unit vectors of the space of the $B_6$ lattice can be computed from the formula $\hat{x}_i = \frac{1}{\sqrt{h}\sqrt{\lambda_i}} \sum_j X_{ji} \frac{2\alpha_j}{(\alpha_j, \alpha_j)}$ where the sixth components of the eigenvectors are taken 1. The orthonormal set of basis vectors of the $B_6$ can be determined as follows

$$\hat{x}_1 = \frac{1}{\sqrt{12}\sqrt{2-\sqrt{2+\sqrt{3}}}}[\sqrt{2-\sqrt{3}}\alpha_1 + \alpha_2 + \sqrt{2}\alpha_3 + \sqrt{3}\alpha_4 + \sqrt{2+\sqrt{3}}\alpha_5 + 2\alpha_6],$$

$$\hat{x}_6 = \frac{1}{\sqrt{12}\sqrt{2+\sqrt{2+\sqrt{3}}}}[-\sqrt{2-\sqrt{3}}\alpha_1 + \alpha_2 - \sqrt{2}\alpha_3 + \sqrt{3}\alpha_4 - \sqrt{2+\sqrt{3}}\alpha_5 + 2\alpha_6],$$

$$\hat{x}_2 = \frac{1}{\sqrt{12}\sqrt{2-\sqrt{2}}}[-\sqrt{2}\alpha_1 - 2\alpha_2 - \sqrt{2}\alpha_3 + \sqrt{2}\alpha_5 + 2\alpha_6],$$

$$\hat{x}_5 = \frac{1}{\sqrt{12}\sqrt{2+\sqrt{2}}}[\sqrt{2}\alpha_1 - 2\alpha_2 + \sqrt{2}\alpha_3 - \sqrt{2}\alpha_5 + 2\alpha_6], \qquad (18)$$

$$\hat{x}_3 = \frac{1}{\sqrt{12}\sqrt{2-\sqrt{2-\sqrt{3}}}}[\sqrt{2+\sqrt{3}}\alpha_1 + \alpha_2 - \sqrt{2}\alpha_3 - \sqrt{3}\alpha_4 + \sqrt{2-\sqrt{3}}\alpha_5 + 2\alpha_6]$$

$$\hat{x}_4 = \frac{1}{\sqrt{12}\sqrt{2+\sqrt{2-\sqrt{3}}}}[-\sqrt{2+\sqrt{3}}\alpha_1 + \alpha_2 + \sqrt{2}\alpha_3 - \sqrt{3}\alpha_4 - \sqrt{2-\sqrt{3}}\alpha_5 + 2\alpha_6].$$

The root vectors of the Coxeter graphs $I_2(\frac{h}{m_i})$ with $m_i = 1, 3, 5$ can be defined as follows:

$$\beta_1 = \sqrt{2}[\sin(\frac{\pi}{24})\hat{x}_1 + \cos(\frac{\pi}{24})\hat{x}_6], \quad \beta_6 = \sqrt{2}[\sin(\frac{\pi}{24})\hat{x}_1 - \cos(\frac{\pi}{24})\hat{x}_6],$$

$$\beta_2 = \sqrt{2}[\sin(\frac{3\pi}{24})\hat{x}_2 + \cos(\frac{3\pi}{24})\hat{x}_5], \quad \beta_5 = \sqrt{2}[\sin(\frac{3\pi}{24})\hat{x}_2 - \cos(\frac{3\pi}{24})\hat{x}_5], \qquad (19)$$

$$\beta_3 = \sqrt{2}[\sin(\frac{5\pi}{24})\hat{x}_3 + \cos(\frac{5\pi}{24})\hat{x}_4], \quad \beta_4 = \sqrt{2}[\sin(\frac{5\pi}{24})\hat{x}_3 - \cos(\frac{5\pi}{24})\hat{x}_4].$$

These sets of vectors define three orthogonal planes determined by the pair of vectors $(\beta_1, \beta_6), (\beta_2, \beta_5), (\beta_3, \beta_4)$ in which the Coxeter element acts like rotations with respective angles $\frac{\pi}{6}, \frac{\pi}{2}, \frac{5\pi}{6}$. Orthogonal projections of a general lattice vector $q = \sum_{i=1}^{5} a_i \omega_i + 2a_6$ onto these planes are given by the following pairs.

The $(\hat{x}_1, \hat{x}_6)$ plane:

$$q_1 = \frac{1}{\sqrt{12}\sqrt{2-\sqrt{2+\sqrt{3}}}}[\sqrt{2-\sqrt{3}}a_1 + a_2 + \sqrt{2}a_3 + \sqrt{3}a_4 + \sqrt{2+\sqrt{3}}a_5 + 2a_6],$$

$$q_6 = \frac{1}{\sqrt{12}\sqrt{2+\sqrt{2+\sqrt{3}}}}[-\sqrt{2-\sqrt{3}}a_1 + a_2 - \sqrt{2}a_3 + \sqrt{3}a_4 - \sqrt{2+\sqrt{3}}a_5 + 2a_6]; \qquad (20)$$

The $(\hat{x}_2, \hat{x}_5)$ plane:

$$q_2 = \frac{1}{\sqrt{12}\sqrt{2-\sqrt{2}}}[-\sqrt{2}a_1 - 2a_2 - \sqrt{2}a_3 + \sqrt{2}a_5 + 2a_6],$$

$$q_5 = \frac{1}{\sqrt{12}\sqrt{2+\sqrt{2}}}[\sqrt{2}a_1 - 2a_2 + \sqrt{2}a_3 - \sqrt{2}a_5 + 2a_6]; \qquad (21)$$

The $(\hat{x}_3, \hat{x}_4)$ plane:

$$q_3 = \frac{1}{\sqrt{12}\sqrt{2-\sqrt{2-\sqrt{3}}}}[\sqrt{2+\sqrt{3}}a_1 + a_2 - \sqrt{2}a_3 - \sqrt{3}a_4 + \sqrt{2-\sqrt{3}}a_5 + 2a_6],$$

$$q_4 = \frac{1}{\sqrt{12}\sqrt{2+\sqrt{2-\sqrt{3}}}}[-\sqrt{2+\sqrt{3}}a_1 + a_2 + \sqrt{2}a_3 - \sqrt{3}a_4 - \sqrt{2-\sqrt{3}}a_5 + 2a_6]. \quad (22)$$

The similar arguments discussed for the $W_a(F_4)$ lattice can apply here. That is, the Voronoi cell $V(0)$ of the $W_a(B_6)$ lattice, the 6D cube represented by the polytope $(0,0,0,0,0,1)_{B_6}$, is projected into the space $E_\perp$ determined by the 4D sub-space defining a window $K$. Projections onto the respective planes $(\hat{x}_1, \hat{x}_6)$ and $(\hat{x}_3, \hat{x}_4)$ lead to the similar results. Figure 6 depicts the projection of the $W_a(B_6)$ lattice onto the plane $(\hat{x}_1, \hat{x}_6)$ which represents a $W_a(B_6)$ quasicrystal with 12-fold symmetry. It is interesting to note that the tiles include rhombi in addition to the usual square and triangle tiles. A recent experimental result (Forster *et al.*, 2013) confirms this kind of tiling.

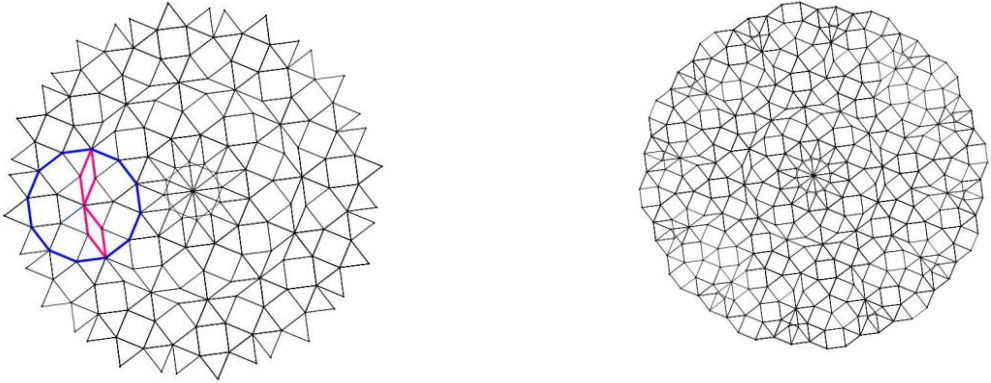

Figure 6. The canonical projection of the lattice $W_a(B_6)$ onto the plane $(\hat{x}_1, \hat{x}_6)$

Let us define the $E_\perp$ space by the union of the planes $(\beta_1, \beta_6)$ and $(\beta_3, \beta_4)$. We project the Voronoi cell into this space to obtain a window. Then we project the lattice points onto the plane $(\hat{x}_2, \hat{x}_5)$. The orthogonal projection leads to a square lattice whose symmetry is the dihedral group $D_4$ of order 8 and its Coxeter-Weyl group is the group $D_4 \approx W(B_2)$. It is another subgroup of the group $W(B_6)$ but not maximal. The projected set is a crystal structure with 4-fold symmetry. A projected section of the lattice is shown in Figure 7.

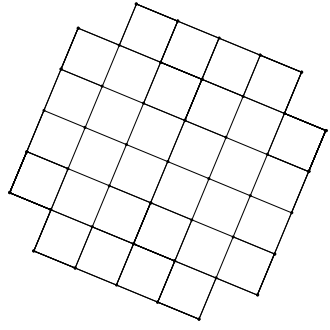

Figure 7. The canonical projection of the lattice $W_a(B_6)$ onto the plane $(\hat{x}_2, \hat{x}_5)$

## 5. Canonical projections of the root and weight lattices of $W_a(E_6)$

The Coxeter-Dynkin diagram of the exceptional group $W(E_6)$ is given in Figure 8. The Corresponding Cartan matrix and its inverse are given as follows:

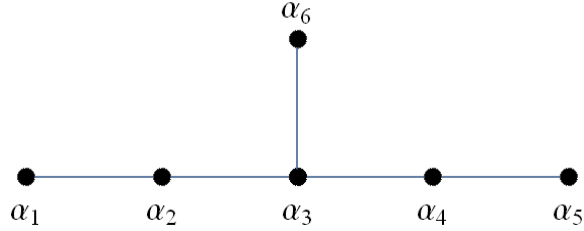

Figure 8. The Coxeter-Dynkin diagram of $E_6$

$$C_{E_6} = \begin{pmatrix} 2 & -1 & 0 & 0 & 0 & 0 \\ -1 & 2 & -1 & 0 & 0 & 0 \\ 0 & -1 & 2 & -1 & 0 & -1 \\ 0 & 0 & -1 & 2 & -1 & 0 \\ 0 & 0 & 0 & -1 & 2 & 0 \\ 0 & 0 & -1 & 0 & 0 & 2 \end{pmatrix}, \quad (C_{E_6})^{-1} = \begin{pmatrix} \frac{4}{3} & \frac{5}{3} & 2 & \frac{4}{3} & \frac{2}{3} & 1 \\ \frac{5}{3} & \frac{10}{3} & 4 & \frac{8}{3} & \frac{4}{3} & 2 \\ 2 & 4 & 6 & 4 & 2 & 3 \\ \frac{4}{3} & \frac{8}{3} & 4 & \frac{10}{3} & \frac{5}{3} & 2 \\ \frac{2}{3} & \frac{4}{3} & 2 & \frac{5}{3} & \frac{4}{3} & 1 \\ 1 & 2 & 3 & 2 & 1 & 2 \end{pmatrix}. \qquad (23)$$

The group generators can be defined by the actions on the simple roots by using the following general formula

$$r_i \Lambda = \Lambda - \frac{2(\Lambda, \alpha_i)}{(\alpha_i, \alpha_i)} \alpha_i, \quad (i = 1, 2, ..., 6). \qquad (24)$$

The order of the Coxeter-Weyl group generated by the 6, $6 \times 6$ matrices is $|W(E_6)| = 2^7 3^4 5 = 51,840$. An arbitrary vector of the root lattice $E_6$ is defined by the vector $p = \sum_{i=1}^{6} b_i \alpha_i$, $b_i \in \mathbf{Z}$. Similarly, a weight vector belonging to the weight lattice $E_6^*$ is given by the vector $q = \sum_{i=1}^{6} a_i \omega_i \equiv (a_1, a_2, ..., a_6)$, $a_i \in \mathbf{Z}$ and the orbit generated by an arbitrary vector is defined by $W(E_6)q \equiv q_{E_6} \equiv (a_1, a_2, ..., a_6)_{E_6}$. If we denote the root and weight lattice vectors by 6-tuples $\vec{b} \equiv (b_1, b_2, ..., b_6)$ and $\vec{a} \equiv (a_1, a_2, ..., a_6)$ respectively then we have the relations $\vec{b}C_{E_6} = \vec{a}$ or $\vec{a}(C_{E_6})^{-1} = \vec{b}$. These relations imply that when $\vec{b}$ has integer components it belongs to the root lattice. It also belongs to the weight lattice since $\vec{a}$ would also takes integer components. However, for $\vec{a}$ with integer components, $\vec{b}$ may not have integer components. Therefore it is clear that the weight lattice $E_6^*$ admits the root lattice $E_6$ as a sublattice. The orthogonal unit vectors for $E_6$ obtained from (5) are given explicitly as follows

$$\hat{x}_1 = \frac{1}{\sqrt{2-\sqrt{2+\sqrt{3}}}\,(2\sqrt{3+\sqrt{3}})}[\alpha_1 + \sqrt{2+\sqrt{3}}\,\alpha_2 + (1+\sqrt{3})\alpha_3 + \sqrt{2+\sqrt{3}}\,\alpha_4 + \alpha_5 + \sqrt{2}\,\alpha_6],$$

$$\hat{x}_6 = \frac{1}{\sqrt{2+\sqrt{2+\sqrt{3}}}\,(2\sqrt{3+\sqrt{3}})}[-\alpha_1 + \sqrt{2+\sqrt{3}}\,\alpha_2 - (1+\sqrt{3})\alpha_3 + \sqrt{2+\sqrt{3}}\,\alpha_4 - \alpha_5 + \sqrt{2}\,\alpha_6],$$

$$\hat{x}_2 = \frac{1}{2}[-\alpha_1 - \alpha_2 + \alpha_4 + \alpha_5],$$

$$\hat{x}_5 = \frac{1}{2\sqrt{3}}[-\alpha_1 + \alpha_2 - \alpha_4 + \alpha_5],$$

$$\hat{x}_3 = \frac{1}{\sqrt{2-\sqrt{2-\sqrt{3}}}\,(2\sqrt{3-\sqrt{3}})}[-\alpha_1 - \sqrt{2-\sqrt{3}}\,\alpha_2 - (1-\sqrt{3})\alpha_3 - \sqrt{2-\sqrt{3}}\,\alpha_4 - \alpha_5 + \sqrt{2}\,\alpha_6],$$

$$\hat{x}_4 = \frac{1}{\sqrt{2+\sqrt{2-\sqrt{3}}}\,(2\sqrt{3-\sqrt{3}})}[\alpha_1 - \sqrt{2-\sqrt{3}}\,\alpha_2 + (1-\sqrt{3})\alpha_3 - \sqrt{2-\sqrt{3}}\,\alpha_4 + \alpha_5 + \sqrt{2}\,\alpha_6].$$

(25)

Using (8) and these unit vectors we define three orthogonal planes with the simple roots $(\beta_1, \beta_6)$, $(\beta_2, \beta_5)$, and $(\beta_3, \beta_4)$ of the graph $I_2(\frac{h}{m_i})$ with $m_1 = 1$, $m_2 = 4$, $m_3 = 5$ respectively. The generators $R_1$ and $R_2$ act like reflection generators on these root spaces and the corresponding Coxeter element acts like rotations by angles $\frac{\pi}{6}, \frac{2\pi}{3}$, and $\frac{5\pi}{6}$ respectively on these orthogonal planes. The dihedral group acting in the planes $(\beta_1, \beta_6)$ and $(\beta_3, \beta_4)$ is non-crystallographic groups however the dihedral group acting in the plane $(\beta_2, \beta_5)$ is a crystallographic group isomorphic to the subgroup $W(A_2)$ of order 6. These properties should be reflected in the projections. Components of an arbitrary weight vector $q = \sum_{i=1}^{6} a_i \omega_i \equiv (a_1, a_2, ..., a_6)$, $a_i \in \mathbf{Z}$ in the respective planes are given as follows.

The plane $(\hat{x}_1, \hat{x}_6)$:

$$q_1 = \frac{1}{\sqrt{2-\sqrt{2+\sqrt{3}}}\,(2\sqrt{3+\sqrt{3}})}[a_1 + \sqrt{2+\sqrt{3}}\,a_2 + (1+\sqrt{3})a_3 + \sqrt{2+\sqrt{3}}\,a_4 + a_5 + \sqrt{2}\,a_6],$$

$$q_6 = \frac{1}{\sqrt{2+\sqrt{2+\sqrt{3}}}\,(2\sqrt{3+\sqrt{3}})}[-a_1 + \sqrt{2+\sqrt{3}}\,a_2 - (1+\sqrt{3})a_3 + \sqrt{2+\sqrt{3}}\,a_4 - a_5 + \sqrt{2}\,a_6];$$

(26)

The plane $(\hat{x}_2, \hat{x}_5)$:

$$q_2 = \frac{1}{2}[-a_1 - a_2 + a_4 + a_5],$$

$$q_5 = \frac{1}{2\sqrt{3}}[-a_1 + a_2 - a_4 + a_5];$$

(27)

The plane $(\hat{x}_3, \hat{x}_4)$:

$$q_3 = \frac{1}{\sqrt{2-\sqrt{2-\sqrt{3}}}\,(2\sqrt{3-\sqrt{3}})}[-a_1 - \sqrt{2-\sqrt{3}}\,a_2 - (1-\sqrt{3})a_3 - \sqrt{2-\sqrt{3}}\,a_4 - a_5 + \sqrt{2}\,a_6],$$

$$q_4 = \frac{1}{\sqrt{2+\sqrt{2-\sqrt{3}}}\,(2\sqrt{3-\sqrt{3}})}[a_1 - \sqrt{2-\sqrt{3}}\,a_2 + (1-\sqrt{3})a_3 - \sqrt{2-\sqrt{3}}\,a_4 + a_5 + \sqrt{2}\,a_6].$$

(28)

Before we study the details of the canonical projection of the $E_6$ lattices we would like to study the projections of certain polytopes of the group $W(E_6)$. The fundamental polytopes are characterized by the orbits $(1,0,0,0,0,0)_{E_6}$, $(0,1,0,0,0,0)_{E_6}$, $(0,0,1,0,0,0)_{E_6}$, $(0,0,0,1,0,0)_{E_6}$, $(0,0,0,0,1,0)_{E_6}$, and $(0,0,0,0,0,1)_{E_6}$. Their vertices are the vectors of the weight lattice. Two particular polytopes belong to both lattices: the root polytope $(0,0,0,0,0,1)_{E_6}$ and the polytope $(0,0,1,0,0,0)_{E_6}$. These two polytopes have a larger symmetry $Aut(E_6) \approx W(E_6):\gamma$ where $\gamma$ represents the Dynkin diagram symmetry of order 2 and (:) means semi-direct product. The characteristic properties of these polytopes, such as, numbers of vertices, edges, 2-facets, 3-facets, 4-facets, and 5-facets (denoted respectively by $N_0$, $N_1$, $N_2$, $N_3$, $N_4$, $N_5$) are listed in Table 3. Their projections onto the plane $(\beta_1, \beta_6)$ are illustrated in Figure 9. The dual of the root polytope $(0,0,0,0,0,1)_{E_6}$ is the union of the polytopes $(1,0,0,0,0,0)_{E_6} \cup (0,0,0,0,1,0)_{E_6}$ and it constitutes the Voronoi cell of the root lattice with $N_0=54$, $N_1=702$, $N_2=2,160$, $N_3=2,160$, $N_4=720$, $N_5=72$. The root polytope as well as its dual has a larger symmetry $Aut(E_6) \approx W(E_6):\gamma$. Their projections are shown in Figure 9.

Table 3. Numbers of facets of some of the $E_6$ polytopes

| Polytope | $N_0$ | $N_1$ | $N_2$ | $N_3$ | $N_4$ | $N_5$ |
|---|---|---|---|---|---|---|
| $(1,0,0,0,0,0)_{E_6}$ | 27 | 216 | 720 | 1,080 | 648 | 99 |
| $(0,1,0,0,0,0)_{E_6}$ | 216 | 2,160 | 5,040 | 4,320 | 1,350 | 126 |
| $(0,0,1,0,0,0)_{E_6}$ | 720 | 6,480 | 10,800 | 6,480 | 1,566 | 126 |
| $(0,0,0,1,0,0)_{E_6}$ | 216 | 2,160 | 5,040 | 4,320 | 1,350 | 126 |
| $(0,0,0,0,1,0)_{E_6}$ | 27 | 216 | 720 | 1,080 | 648 | 99 |
| $(0,0,0,0,0,1)_{E_6}$ | 72 | 720 | 2,160 | 2,160 | 702 | 54 |

These numbers satisfy the Euler's equation $N_0 - N_1 + N_2 - N_3 + N_4 - N_5 = 0$.

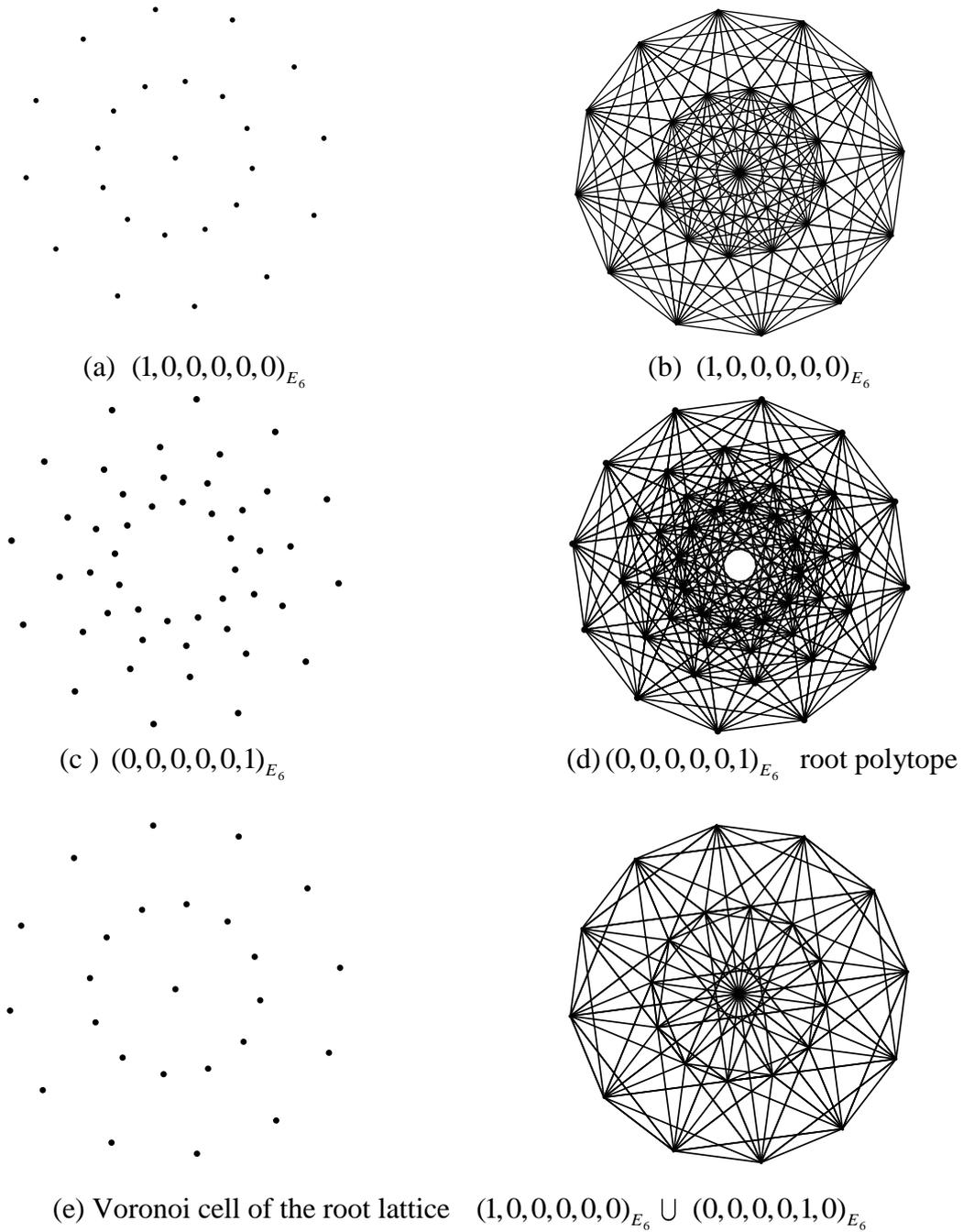

(a) $(1,0,0,0,0,0)_{E_6}$      (b) $(1,0,0,0,0,0)_{E_6}$

(c) $(0,0,0,0,0,1)_{E_6}$      (d) $(0,0,0,0,0,1)_{E_6}$ root polytope

(e) Voronoi cell of the root lattice    $(1,0,0,0,0,0)_{E_6} \cup (0,0,0,0,1,0)_{E_6}$

Figure 9. Projections of some of $E_6$ polytopes onto the Coxeter plane

Now we can choose any one of these planes as $E_{\parallel}$ and the rest of the four dimensional space as $E_{\perp}$. Let us choose the plane $(\beta_1, \beta_6)$ as $E_{\parallel}$ and the rest as $E_{\perp}$. As we have explained in Section 2 the canonical projection is carried out by projecting the Voronoi cell $V(0)$ of the root/weight lattice into the space $E_{\perp}$ to obtain a window $K$. The Voronoi cell of the weight lattice $E_6^*$ (Conway & Sloane, 1988; Worley, 1987; Pervin, 2013) is a polytope which can be specified as $\frac{1}{3}(0,0,1,0,0,0)_{E_6}$. It has

$$N_0 = 720 \text{ vertices}, N_1 = 6,480 \text{ edges}, N_2 = 10,800 \text{ triangular faces}, \tag{29}$$

$N_3 = 6,480$ 3-facets (2,160 tetrahedra+4,320 octahedra), $N_4 = 1,566$ 4-facets,

$N_5 = 126$ 5-facets.

The 720 vertices of the polytope are determined by the action of the group elements on the vector $\frac{1}{3}(0,0,1,0,0,0)$ and using the formula

$$R_0 = \frac{1}{3}(q_2^2 + q_3^2 + q_4^2 + q_5^2)^{\frac{1}{2}} \tag{30}$$

one obtains the window **K** in the form of a 3-sphere. Now we project the lattice vectors into this window using (26-28). Next step is to project the lattice vectors which fall in the window **K** onto the plane $(\beta_1, \beta_6)$ by using the components of the vectors given in (26). We illustrate the projection of the root lattice in Figure 10. They are the same point distributions in which (b) represents the tiling with minimal distance and (c) represents a partial tiling which is exactly the same tiling obtained from the projection of $F_4$ lattice that is a sublattice in the root lattice $E_6$.

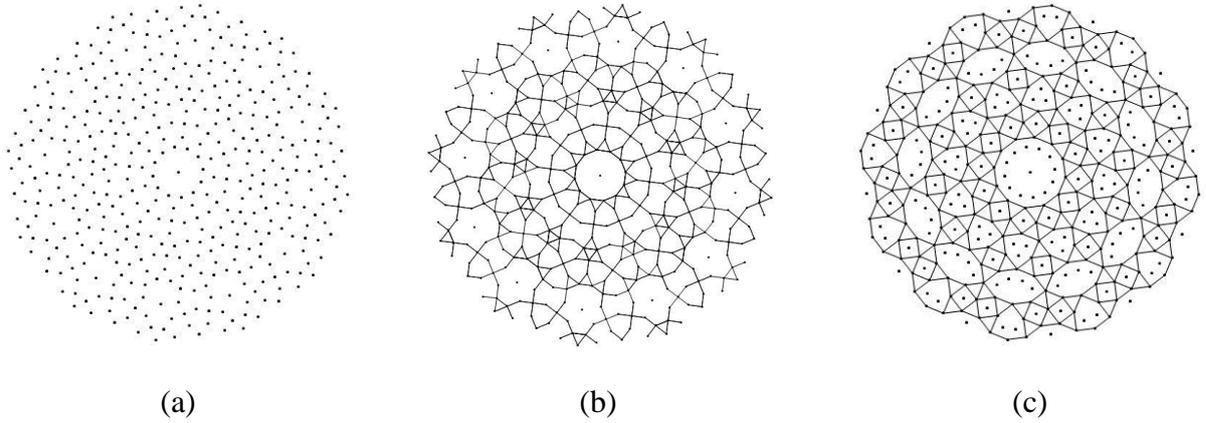

(a)          (b)          (c)

Figure 10. Root lattice $E_6$

The projection of the weight lattice is depicted in Figure 11. We note the fact that the similar distributions are obtained if the window **K** is chosen to be the 3-sphere $R_0 = \frac{1}{3}(q_1^2 + q_2^2 + q_5^2 + q_6^2)^{\frac{1}{2}}$ and the projection of the lattice is made onto the plane $(\beta_3, \beta_4)$.

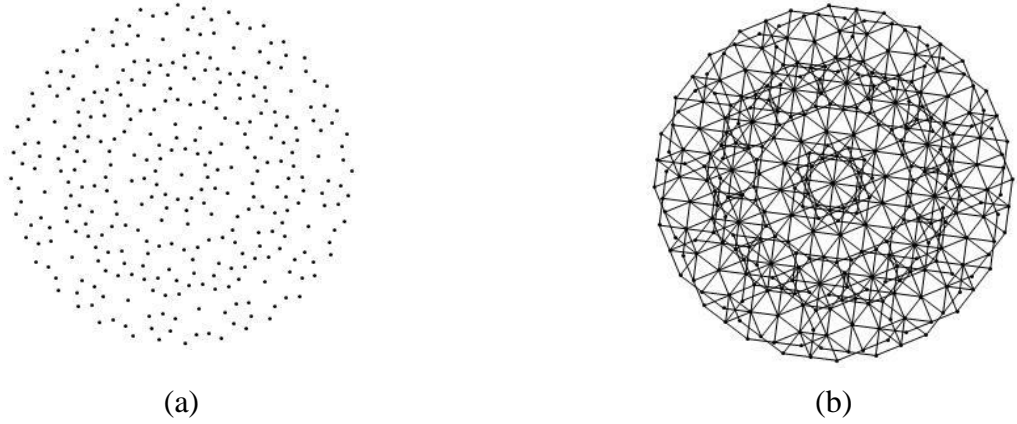

(a)                      (b)

Figure 11. Canonical projection of the weight lattice of $E_6$ onto the plane $(\beta_1, \beta_6)$

Now let us define the $E_\parallel$ plane by the pair of vectors $(\beta_2, \beta_5)$. The representation of the Coxeter group in the plane $(\beta_2, \beta_5)$ is such that $(R_1 R_2)^3 = 1$. Therefore the dihedral group is represented by the group $D_3 \approx S_3 \approx W(A_2)$ which is not maximal. Consequently, the 4-dimensional Euclidean sub-space $E_\perp$ is determined by the orthogonal planes $(\beta_1, \beta_6)$ and $(\beta_3, \beta_4)$. This time we consider only the weight lattice projection in which the window $K$ is determined by the projection of the Voronoi cell $\frac{1}{3}(0,0,1,0,0,0)_{E_6}$ into the space $E_\perp$. After repeating the previous procedure we project the lattice points into the plane determined by the vectors $(\beta_2, \beta_5)$ and obtain the planar lattice as shown in Figure 12. It is a honeycomb lattice as expected because the projection has been made onto a plane where the crystallographic subgroup $W(A_2) \subset W(E_6)$ acts. It is interesting to observe that the honeycomb lattice exits in nature as the graphene sheet.

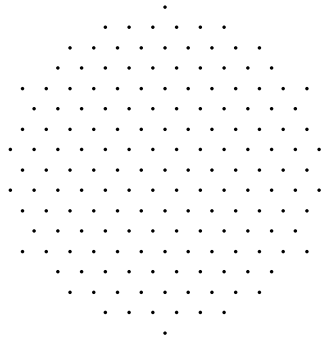

Figure 12. The honeycomb lattice obtained by projection of $E_6$ lattice

## 6. Conclusion

This paper has introduced a general technique applicable to the projections of all higher dimensional lattices generated by the affine Coxeter-Weyl groups. It was shown that the eigenvalues, the corresponding eigenvectors, and the simple roots of the Cartan matrix (Gram matrix) play an important role in the projection technique. The maximal dihedral subgroup of the Coxeter-Weyl group can be generated by two generators whose product defines the Coxeter element of the Coxeter-Weyl group. It was noted that the Coxeter group acts, in the lattice space spanned by the orthogonal unit vectors, as a reducible representation of the dihedral group in the block diagonal form of $2\times 2$ irreducible representations and/or as $1\times 1$ matrix depending on the rank of the Coxeter-Weyl group. If the dihedral subgroup acting on the principal plane (Coxeter plane) is non-crystallographic the point distribution represents a quasicrystal with $h$-fold symmetry where $h$ is the Coxeter number; otherwise it represents a crystallographic point distribution such as square and/or hexagonal lattices. We have studied examples of 12-fold symmetry by projection of the lattices $W_a(F_4)$, $W_a(B_6)$, and $W_a(E_6)$. It turned out that the structures obtained by the projections of the lattices $W_a(F_4)$ and $W_a(B_6)$ are compatible with the 12-fold symmetric quasicrystal structure observed in Ni-Cr particles (Ishimasa *et al*., 1985) and in dodecagonal quasicrystal formation $BaTiO_3$ (Forster *et al*., 2013) respectively. The projection technique we introduced can be applied to any lattice described by the affine Coxeter group.